\documentclass[%
 reprint,
 amsmath,amssymb,
 aps,
]{revtex4-2}

\usepackage{graphicx}
\usepackage{dcolumn}
\usepackage{multirow}
\usepackage{bm}

\begin{document}


\title{On the interpretation of non-resonant phenomena at colliders}

\author{Miguel G. Folgado}
\email{migarfol@ific.uv.es}
\affiliation{Instituto de F\'isica Corpuscular (IFIC), Universidad de Valencia-CSIC, E-46980, Valencia, Spain}

\author{Veronica Sanz}%
 \email{veronica.sanz@uv.es}
\affiliation{Department of Physics and Astronomy, University of Sussex, Brighton BN1 9QH, UK}

\affiliation{Instituto de F\'isica Corpuscular (IFIC), Universidad de Valencia-CSIC, E-46980, Valencia, Spain}

\date{\today}


\begin{abstract}
With null results in resonance searches at the LHC, the physics potential focus is now shifting towards the interpretation of non-resonant phenomena. An example of such shift is the increased popularity of the EFT programme. We can embark on such programme owing to good integrated luminosity and an excellent understanding of the detectors, which will allow these searches to become more intense as the LHC continues. In this paper we provide a framework to perform this interpretation in terms of a diverse set of scenarios, including {\it 1.) } generic heavy new physics described at low energies in terms of a derivative expansion, such as in the EFT approach, {\it 2.)} very light particles with derivative couplings, such as axions or other light pseudo-Goldstone bosons, and {\it 3.)} the effect of a quasi-continuum of resonances, which can come from a number of strongly-coupled theories, extra-dimensional models, clockwork set-ups and their deconstructed cousins.  These scenarios are not equivalent despite all non-resonant, although the matching among some of them is possible and we provide it in this paper. 
\end{abstract}
\maketitle

\section{\label{sec:intro} Why non-resonant phenomena}
 
 At the Large Hadron Collider (LHC), a tremendous effort has been devoted to searches for new states, which in their simplest form would manifest as resonances, an excess of events in a narrow kinematic range. Alas, {\it direct searches} have provided no fruit so far, diminishing hopes that the LHC would discover new particles connected to the electroweak scale. 
 
\begin{figure}[h!]
    \centering
    \includegraphics[scale=0.1]{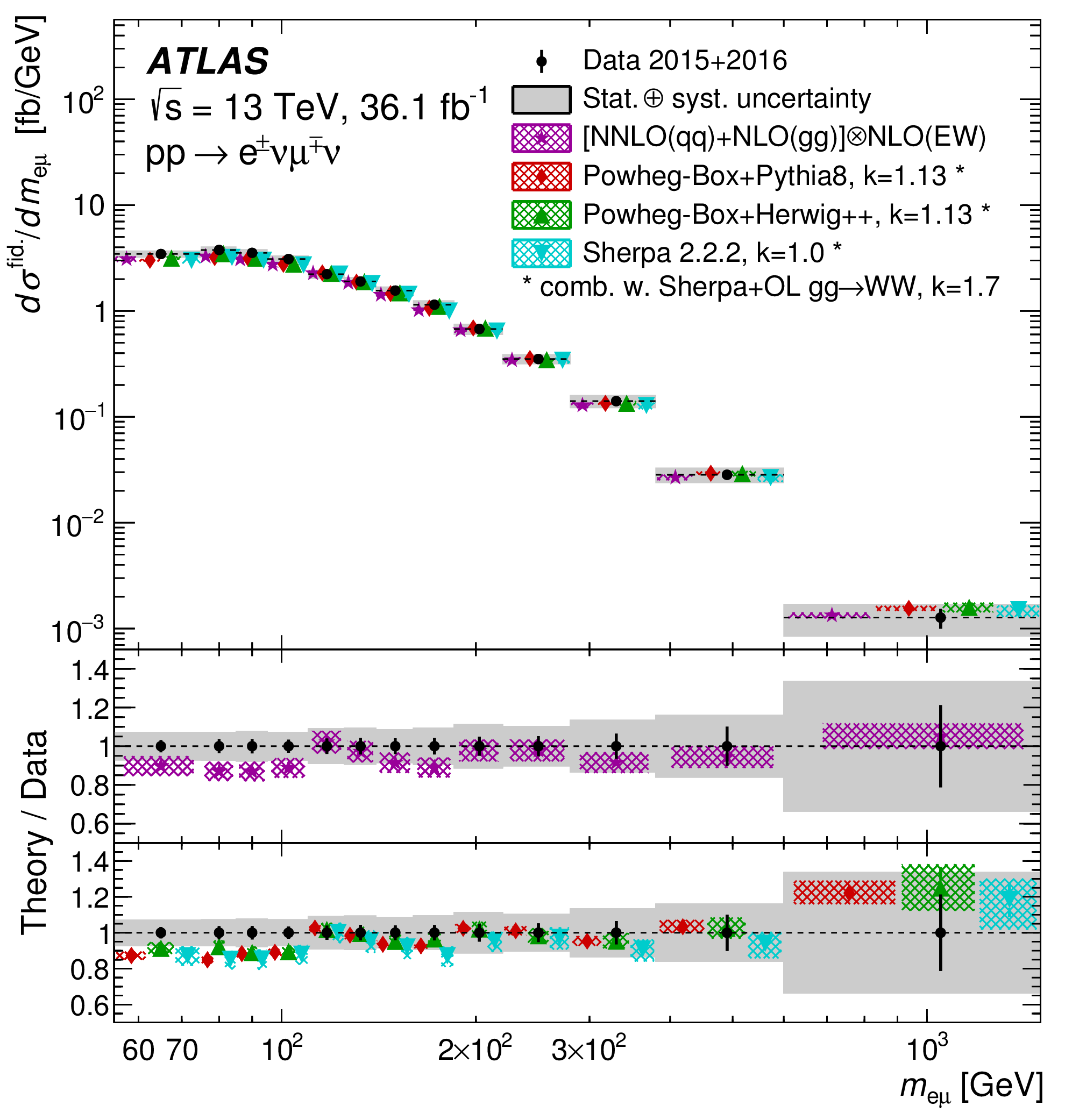}
    \caption{An example of a precise determination of a kinematic distribution (invariant mass of the two leptons in $W^+W^-$ production) and its comparison with different theoretical calculations and showering tools. Figure from Ref.~\cite{Aaboud:2019nkz}.}
    \label{fig:exampleWW}
\end{figure}

 At the same time, LHC experiments have continued improving their understanding of the data, and enabled a unprecedented precise characterization of Standard Model (SM) particles. Particularly impressive is the evolution of the Higgs characterization, moving from  discovery to precision measurements in a few years.  
In Fig.~\ref{fig:exampleWW} we can appreciate the level of effort made to contrast precise measurements of the LHC experiments and our best theoretical understanding.

 The lack of discoveries is certainly a frustrating aspect of the LHC results, but with improved precision a new opportunity opens up. The same SM particles whose characterization is so carefully accomplished could provide hints for new phenomena beyond the SM.  Among all the SM particles, high hopes  are placed on heavy particles, more closely connected to the electroweak symmetry breaking sector: the Higgs particle, massive vector bosons, and third-generation fermions. Were their properties deviate from the SM predictions,  we might be able to trace back these deviations to new physics. This {\it indirect} route may be the key to enter new sectors in Nature.  
 
 These two approaches of searching for new physics, {\it direct} search for new states and {\it indirect} search for new effects in known SM particles, complement each other. There is a wide range of new theories of Nature which at the LHC energies do no show up as narrow resonances, and for those the best way to find them is by looking at excesses in tails of kinematic distributions. 
 
The aim of this paper is to provide a framework to interpret non-resonant phenomena in terms of a diverse set of scenarios. We will provide a number of well-motivated models which, in some area of their parameter space, leads to no resonances but excesses in tails.  

The first paradigm we will discuss is the Effective Field Theory (EFT) approach, suitable for theories with resonances  too heavy to be accessed directly at the LHC, and only showing up as virtual effects. 

The EFT is a well-known and nowadays a rather standard way of organising non-resonant searches, but it is certainly not the only one. There are other interesting theories which exhibit the same experimental signatures. One class of such models is represented by many close-by resonances forming a quasi-continuum in $\sqrt{\hat s}$, the parton energy. This type of behaviour does appear in many extensions of SM which predict towers of resonances with the same quantum numbers but increasing mass. Were these towers broad/close-by, they would lead to non-resonant tails. Typical examples of this type of phenomena appear in theories with new strong interactions, extra-dimensions or clockwork theories, to name a few.  

Another distinct set of theories is  which could be only be discovered via non-resonant searches are scenarios with light particles, so light that usual trigger cuts would prevent us from reaching their resonant kinematic regions, but could still be discovered by their off-shell production. If the coupling of these light states to SM particles were derivative, the off-shellness could be compensated by the natural increase of the cross-section. A paradigmatic example of this kind of scenarios is an axion or axion-like particle coupled to SM gauge bosons, see Ref.~\cite{Gavela:2019cmq} for a discussion of this behaviour.

The paper is organized as follows. In Sec.~\ref{sec:models} we discuss the models which can provide a basis for experimental interpretation. In Sec.~\ref{sec:pheno}, we discuss parton-level properties of the cross sections and provide a dictionary to match parameters among the models which share common experimental tails. We also discuss how these parton-level results generalise to a total production cross section. Finally, in Sec.~\ref{sec:conclusions} we summarise our findings.  
 
\section{\label{sec:models} Models of non-resonant behaviour}

\begin{figure}[h!]
    \centering
    \includegraphics[scale=0.23]{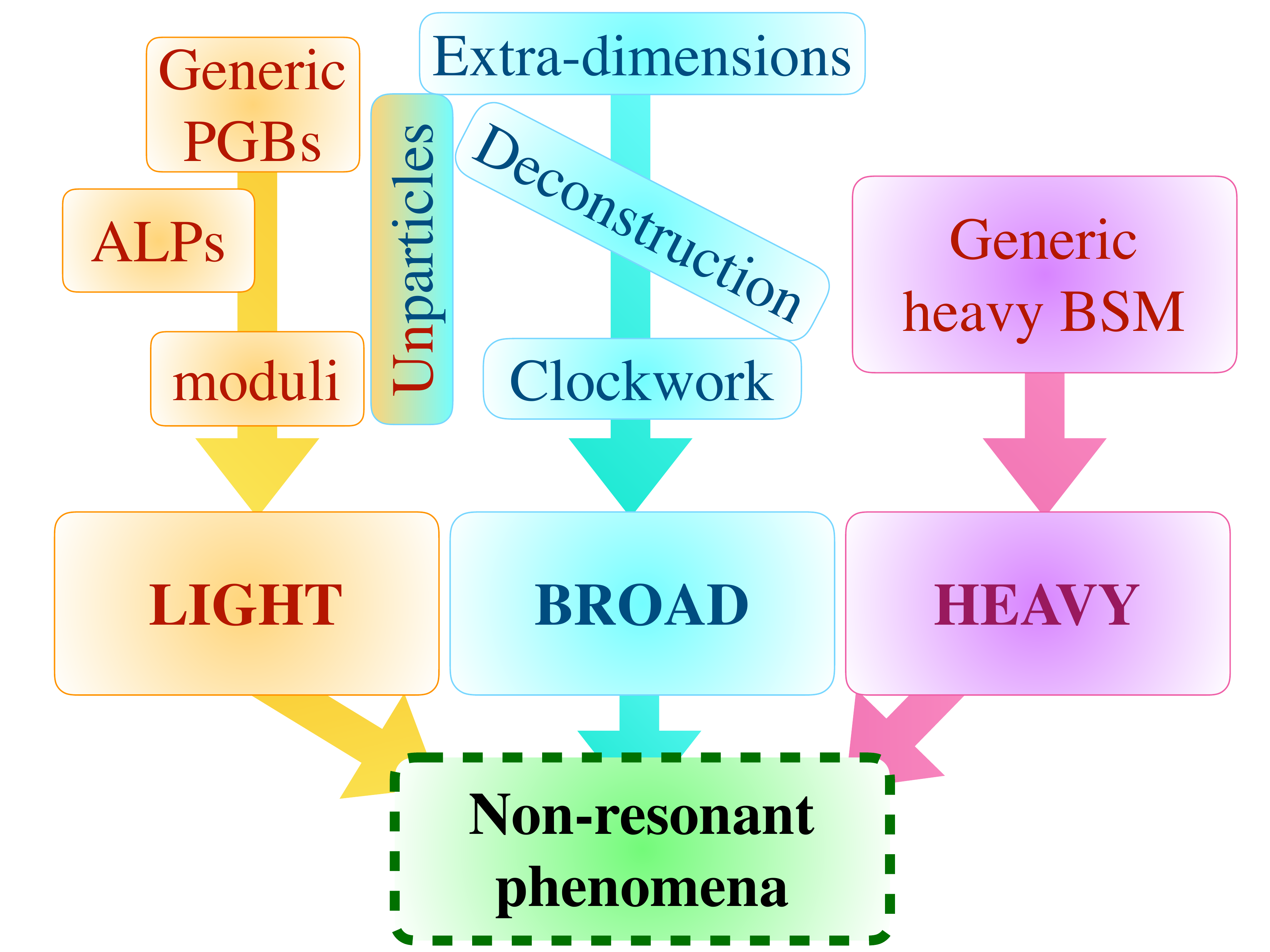}
    \caption{Schematic view of models of non-resonant behaviour.}
    \label{fig:mindmap}
\end{figure}

In this section we describe New Physics scenarios which lead to non-resonant behaviour, as well as relations among them. Broadly speaking these scenarios fall into three categories: 

\begin{enumerate}
\item Heavy but narrow new states, due to some new sector well above the electroweak scale
\item Broad/nearby states at the electroweak scale
\item Light states, with mass well below electroweak's, and with derivative couplings 
\end{enumerate}

Many models of new physics, at some scale or in some range of their parameter space could show up at colliders as non-resonant excesses. In the rest of the section, we provide some context for some of the most popular models in each category. In the next section, we will provide a dictionary to relate scenarios of broad/nearby resonances, hence unifying their interpretation.  

\subsection{Heavy physics: the Effective Field Theory (EFT) case}
We start with the paradigmatic example of EFTs. In this framework one studies theories which predict a new sector of resonances coupled to the SM, with masses higher than the energies one is able to probe with current experiments. These new resonances will not be produced as final on-shell  states, but can influence the behaviour of SM particles through their virtual effects. 

\begin{figure}[h!]
    \centering
    \includegraphics[scale=0.23]{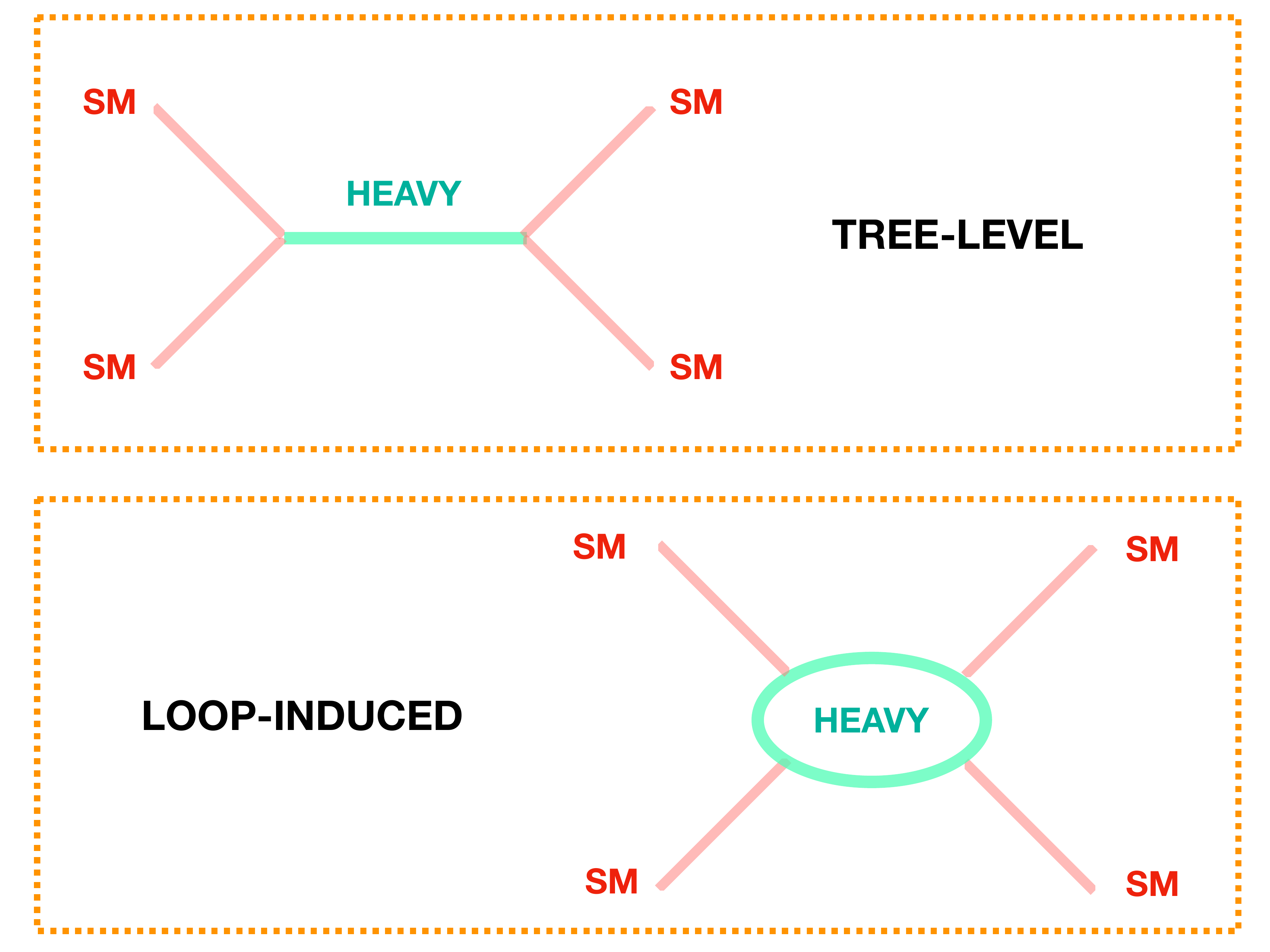}
    \caption{Sketch of the UV origin of EFT effects.}
    \label{fig:EFT}
\end{figure}

One could consider the  effect of new resonances on a scattering or decay process via tree-level (Fig.~\ref{fig:EFT} top) or loop (e.g. Fig.~\ref{fig:EFT} bottom) exchange of the heavy states. As the resonances are heavy, $M \gg \sqrt{\hat s}$, one can understand the impact of the new sector as an expansion in energy over mass with various Lorentz structures involving just SM states as dynamical. 
For example, let us consider classes of models which affect the electroweak symmetry breaking sector of the SM. Their presence would modify the interactions of the Higgs and massive gauge bosons, e.g. via new types of interactions described by new operators in the Lagrangian such as 
\begin{equation}
\bar c_{HW} \frac{2 \, g}{m_W^2} \, \left( D^\mu \, H^\dagger \, (T_2)_k \, D^\nu \, H\right) \, W_{\mu\nu}^k
    \label{eq:EFT1}
\end{equation}
where $H$ is the Higgs doublet and $W_{\mu\nu}$ is the $SU(2)_L$ field strength. 
These new type of interactions are formally non-renormalisable and grow with the energy of the scattering process. At the level of Higgs-vector boson interactions, they will modify the coupling structure as~\cite{Alloul:2013naa} 
\begin{eqnarray}
    & &[h(p_1), \, W^\mu(p_2), \, W^\nu(p_3)] = i \, g \, m_W [ \eta^{\mu\nu} +  \nonumber \\
    & & \frac{\bar c_{HW}}{m_W^2} \, (2 p_2.p_3+(p_2^2+p_3^2)-2 p_2^\nu \, p_3^\mu - (p_2^\mu p_2^\nu+p_3^\mu p_3^\nu) ]
\end{eqnarray}

These new types of interactions grow with energy and exhibit different angular dependence than the simple SM $\eta^{\mu\nu}$ structure. They will lead to excesses in high momentum tails of production of SM gauge and Higgs bosons, such as $W^+ W^-$ distributions shown in Fig.~\ref{fig:exampleWW}. The behaviour of this tail, the rate at which it grows with energy, will be dictated by the form of the operator in Eq.~\ref{eq:EFT1} and the value of its coefficient $\bar c_{HW}$, but their interpretation in terms of new sectors of Nature could be very different. 

For example, in Ref.~\cite{Gorbahn:2015gxa} we computed three possible interpretations of these effects in terms of new scalars coupled to the electroweak sector,
\begin{eqnarray}
    \bar c_{HW} & \simeq &  \, \lambda_{hS}^2 \frac{v_s^2 v^2}{m_S^4} \textrm{\hspace{1cm} \bf ($SU(2)_L$ Singlet)} \nonumber\\
     & \simeq &  \, -\frac{2 \lambda_{hH}}{192\, \pi^2} \, \frac{m_W^2}{m_{H_2}^2}\, \textrm{ \bf ($SU(2)_L$ Doublet)} \nonumber\\
      & \simeq &  \, -\frac{c_{CFT}}{4}\, \frac{m_h^2 \, v^2}{f^2 \, m_r^2} \, \textrm{ \bf (Radion/Dilaton)}
\end{eqnarray}
where $m_{S,H_2,r}$ are the singlet, doublet and radion/dilaton masses, $\lambda_{hS, hH}$ are the singlet/doublet-Higgs couplings, $v_S$ is the singlet vev, and $c_{CFT}$ is a parameter of the new quasi-conformal sector. See Ref.~\cite{Gorbahn:2015gxa} for details, as well as Ref.~\cite{Masso:2012eq} to understand how  the singlet expression has been obtained.  

Despite sharing a common experimental search (same cuts and statistical procedure), the limit on the value of $\bar c_{HW}$ would translate into very different limits for the three scenarios discussed above. 

\subsection{Light new physics: axions and other pseudo-Goldstone bosons}

Since Goldstone obtained his theorem, we know that light degrees of freedom (Goldstone bosons) are  a manifestation of a global symmetry which is spontaneously broken by some UV dynamics, e.g. confinement from a strong gauge sector. We know this description to be very predictive in the low-energy sector of QCD ($\chi$PT), and it is expected that the same mechanism  re-appears in other sectors of Nature. For example, this mechanism is called for as a solution to the strong CP-problem of QCD, leading to the hypothesis that a new light state, called axion, would couple to SM particles. 

And for many new physics scenarios we design, especially if they involve new strong interactions, these (pseudo-)Goldstone bosons are present and couple to SM particles via interactions which show growth with energy, similar to the EFT case. 

Indeed, if we denote the new light state as $a$, and assumed for simplicity to have pseudo-scalar quantum numbers, the interaction with the SM could look like 
\begin{equation}
    \frac{\alpha}{f_a} \, a \, F^{\mu\nu} \tilde F_{\mu\nu} \ ,
\end{equation}
where $F_{\mu\nu}$ denotes the field strength of any of the SM interactions and $\alpha$ the strength. This interaction is dimension-five, grows with $\hat s$ and is suppressed by the scale $f_a$, the breaking scale of the UV global symmetry which led to the existence of $a$. Apart from couplings to gauge bosons, these pseudo-Goldstone bosons could couple to fermions, with a strength typically suppressed by the fermion mass. 

Experimentally, distinguishing a genuinely EFT excess from an axion-like excess requires an extended understanding of the tail's behaviour. Hence both interpretations should be made side-by-side until an excess is found, and the resolution of this excess allows to disentangle both hypothesis.

\subsection{Broad resonances: Models of towers of states} 

Apart from heavy and light resonances, one could also interpret tails in energy distributions in terms very close-by  resonances, so close that they form an almost continuum. There are many models which could produce such behaviour, from towers of resonances with the same quantum numbers to quasi-conformal theories. In this section we will propose some benchmarks for interpretation and in Section~\ref{sec:pheno} provide a dictionary to relate them.

\subsubsection{Warped Extra-Dimensions} 
The EFT framework is very popular for many reasons, including its generality and ability to be improved order-by-order in an expansion of momenta over the heavy masses. Yet, EFTs are based on the assumption that the new sector is at a scale well separated from the electroweak scale, hence losing ground in terms of understanding the origin of the electroweak sector, including the {\it hierarchy problem}.

This problem has taken up a lot of the community's minds for a long time, and accordingly been rephrased in many ways. One particularly compelling view to solve the hierarchy problem comes from linking gravitational interactions with the SM, in particular assuming that the scale of strong gravitational effects is much lower than the {\it perceived} Planck scale, all the way down to the TeV scale, close to the electroweak scale. This can be achieved by postulating the existence of new, small dimensions of space. 

\begin{figure}[htbp]
\centering
\includegraphics[width=60mm]{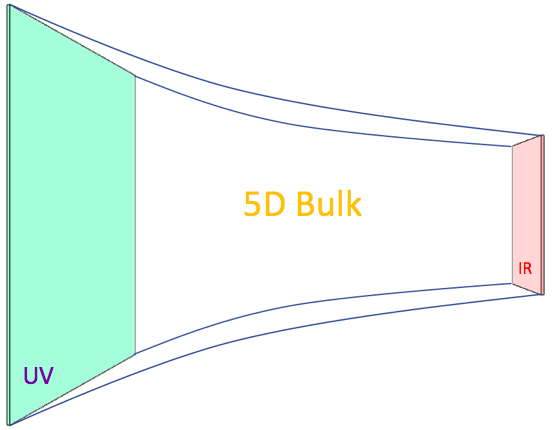}
\caption{Schematic view of the RS geometry. The warp factor decreases exponentially along the bulk.}
 \label{fig:diagramaRS}
\end{figure}

These new theories of extra-dimensions and strong gravity at the TeV scale could take many forms, encoded in the choice for space-time geometry and localisation of the fields inside the extra-dimension. A very well-studied and successful framework is the so-called Randall-Sundrum (RS) model~\cite{Randall:1999ee}, or simply Warped Extra-Dimensions. It is based on a set-up with a single new dimension of space with a large curvature, and a non-factorisable metric
\begin{equation}
    ds^2 = e^{-2 k r_c |y|} \eta_{\mu \nu} dx^\mu dx^\nu - r_c^2 dy^2 \ ,
    \label{eq:RSmetric}
\end{equation}
where $k$ is the 5D curvature, $2 \pi r_c$ is the size of the extra-dimension and $y$ is the coordinate in the 5th dimension.   The RS models is built on a slice of the 5D space-time located between two 4D branes, which are typically located at $y=0$ (UV-brane) and $y=\pi$ (IR-brane),  see Fig.~\ref{fig:diagramaRS}. 
The combination $e^{-2 k r_c |y|}$ is the {\it warp factor} and affects 4D slices of space-time in different ways, exponentially different as we move along the bulk. In the classical 4D space-time the gravitational fundamental interaction is the Planck mass, $\bar{M}_p = M_p/\sqrt{8\pi} = 2.435 \times 10^{18}$ GeV. The effect of the new extra-dimension is to modify the fundamental scale 
\begin{equation}
    \bar{M}_p^2 = \frac{M_5^3}{k}(1 - e^{-2k\pi r_c}) \ , 
    \label{eq:M5}
\end{equation}
and to lower the scale of strong gravitational phenomena (like Black Hole formation) to the TeV scale near the IR brane.

Warped models come in different varieties depending on particle localizations as well as the exact form of the geometry. Nevertheless, we generically expect the existence of towers of massive spin-2 resonances (Kaluza-Klenin gravitons) as a consequence of compactifying higher-dimensional gravity. Indeed, one can expand fluctuations around the 4D projection of the 5D metric to obtain
\begin{equation}
    G^{(5)}_{\mu \nu} =e^{-2\sigma}(\eta_{\mu \nu} + \frac{2}{M_5^{3/2}}h_{\mu \nu}),
    \label{eq:lineargravity}
\end{equation}
where $h_{\mu \nu}$ can then be interpreted as a 5D  graviton. After compactification, this 5D field can be written as a massless graviton and an infinite KK-tower of massive gravitons. The effective scale interaction of the  KK-gravitons depends on the location in bulk of field it interacts with. Fields near the IR couple to the  KK-tower with scale $\Lambda\sim$ TeV, namely
\begin{equation}
    \mathcal{L} = - \frac{1}{\Lambda} \sum_{n=1}^\infty h^n_{\mu \nu}(x)T^{\mu \nu}(x),
    \label{eq:lagrangiano}
\end{equation}
where $T^{\mu \nu}$ is the SM energy-momentum tensor while $h^n_{\mu \nu}$ is the $n$-th KK-graviton.

The spectrum of the KK-tower, how broad and spaced the states are, depends on the exact geometry, whether it takes the simple form of RS (Eq.~\ref{eq:RSmetric}) or something more complex. For example, if during compactification there were background fields in the bulk, the effective geometry could approximate RS near the UV brane, but deviate substantially near the IR~\cite{Hirn:2005nr,Hirn:2006nt} and hence drastically changing the phenomenology~\footnote{See Refs.~\cite{Hirn:2005vk,Hirn:2007we,Hirn:2008tc} for examples on how background fields can change the spectrum of KK states }. 

In the simple case of RS, one can find the KK spectrum by solving the Einstein equation and the equation of motion for the KK-graviton,
\begin{equation}
    m_n = k x_n e^{-k\pi r_c},
    \label{eq:masas}
\end{equation}
where $x_n$ are the roots of the $J_1$ Bessel function.

Finally, in RS or any other model of extra-dimensions, keeping a finite size for an extra-dimension (in our case, at $y = \pi r_c$) is not trivial, and usually relies on introducing new degrees of freedom. A popular mechanism of stabilisation is the Goldberger-Wise~\cite{Goldberger:1999uk}, which calls on a new 5D scalar that mixes with the graviscalar ($G_{55}$), whose vev controls $r_c$,  dynamically stabilizing the size of the extra-dimension. This field can be written as a KK-tower, and its light zero-mode is usually called {\it radion} (already mentioned in the EFT section). Its mass is a free parameter, and interacts with SM fields via the trace of the stress tensor, $-\frac{1}{\sqrt{6} \, \Lambda} \, r \, T^\mu_\mu$.
As the strength of interactions is usually weaker than that of KK-gravitons in the rest of the paper we only consider KK-gravitons, but the discussion could be extended to the radion.

\subsubsection{Deconstruction}
\begin{figure}[htbp]
\centering
\includegraphics[width=80mm]{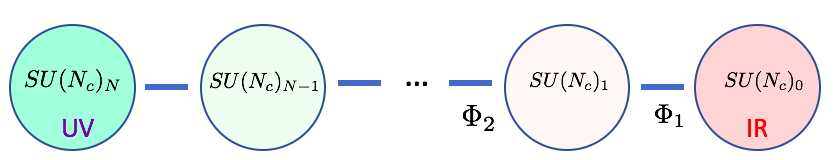}
\caption{$SU(N_c)$ lattice representation of Deconstruction.}
 \label{fig:fiagramaDeco}
\end{figure}
 Towers of resonances can appear in many ways, not just as a consequence of the compactification of extra-dimensions. In this section we discuss the first alternative interpretation of non-resonant phenomena produced by broad/nearby resonances. Our set-up is {\it deconstruction}~\cite{ArkaniHamed:2001ca}, namely the idea that certain types of gauge theories could resemble extra-dimensions as long as they provide a {\it latticised} version of a real (continuous) space coordinate, see Fig.~\ref{fig:fiagramaDeco}. The origin of deconstructed models goes far back to old models of {\it mooses}~\cite{Georgi:1985hf}, series of new confining gauge sectors which at low energies exhibit interesting properties.
 
The simplest deconstruction framework assumes $N + 1$ copies of $SU(N_c)$, where $N_c$ refers to a new QCD-type colour, and $N$ Higgs-like fields with a new VEV ($v$). The Lagrangian of these new fields can be written as
\begin{equation}
\mathcal{L} = -\frac{1}{4} \sum_{i=1}^N F^a_{i \mu \nu}F^{i \mu \nu a} + \sum_{i=1}^N D_\mu \Phi^{\dagger}_i D^\mu \Phi_i,
\end{equation}
where 
\begin{equation}
D_{\mu} = \partial_\mu + ig \sum_{i=1}^N A_{i \mu}^a T^a_i, 
\end{equation}
and one typically one assumes all the $\Phi_i$ fields couple with the same coupling ($g$).

These VEVs break  $N+1$ $SU(N_c)$ symmetry to the diagonal $SU(N_c)$ gauge group. The mass-matrix for the gauge fields is given by
\begin{equation}
\mathcal{M} = \frac{1}{2} g v
\begin{pmatrix}
1 & -1 & 0 & \cdot \cdot \cdot & 0 & 0\\
-1 & 2 & -1 & \cdot \cdot \cdot & 0 & 0\\
0 & -1 & 2 & \cdot \cdot \cdot & 0 & 0\\
\vdots & \vdots & \vdots & \cdot \cdot \cdot & \vdots & \vdots\\
1 & 0 & 0 & \cdot \cdot \cdot & 2 & -1\\
1 & -1 & 0 & \cdot \cdot \cdot & -1 & 1
\end{pmatrix}\ \ ,
\nonumber
\end{equation}
and the gauge fields mass eigenstates can be obtained by diagonalizing this matrix. The mass for the $n$-th gauge field can be written as
\begin{equation}
M_n = 2 g v \sin \left( \frac{n\pi}{2(N + 1)} \right),
\end{equation}
which for $n \ll N$ approximates to 
\begin{equation}
M_n \approx \frac{g v \pi n}{N+1}.
\end{equation}

\subsubsection{ClockWork/Linear Dilaton}

Here we discuss another example of models with limiting behaviour of broad/nearby resonances, {\it clockwork} models. They were born as a toy set-up able to encapsulate some of the known phenomenology of extra-dimensions and deconstruction, but allowing more flexibility of parameter choices. In particular, we will discuss one particular version of Clockwork, the so-called continuum Clockwork/Linear Dilaton (CW/LD) model, which was proposed in Ref.~\cite{Giudice:2016yja} and whose non-resonant phenomenology has been studied in Ref.~\cite{Giudice:2017fmj}. 

\begin{figure}[htbp]
\centering
\includegraphics[width=60mm]{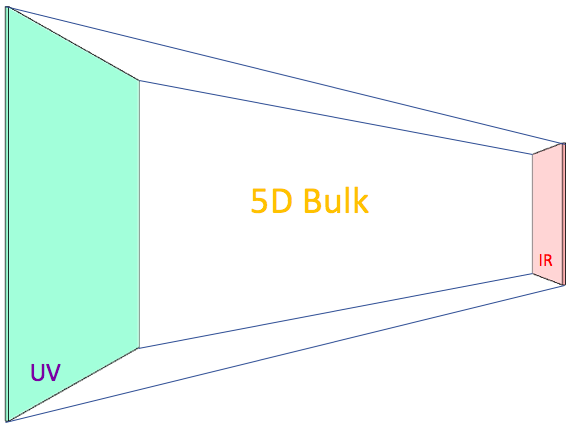}
\caption{Schematic view of the ClockWork/Linear Dilaton model.}
 \label{fig:fiagramaCW}
\end{figure}

The metric that describes this scenario can be written as
\begin{equation}
     ds^2 = e^{4/3 k r_c |y|} \left( \eta_{\mu \nu} dx^\mu dx^\nu - r_c^2 dy^2  \right) \ ,
    \label{eq:CWmetric}
\end{equation}
where the  parameter $k$, called {\it clockwork spring}, represents the curvature along the 5th-dimension and $r_c$, as in RS case, represents its size. 

There are many similarities with RS framework, the compact 5D space-time also sandwiched between the UV-brane (at $y=0$) and IR-brane (at $y=\pi$). The SM could be localised in the 5D bulk, but in the minimal scenario it is confined on the IR-brane. The relation between the fundamental 4D and 5D scales is given by 
\begin{equation}
     \bar{M}_p^2 = \frac{M_5^3}{k} \left( e^{2\pi k r_c} - 1  \right).
    \label{eq:M5CW}
\end{equation}
In RS $M_5 \sim \bar{M}_p$ and the interaction scale in the IR-brane is order TeV while in CW/LD $M_5$ is directly order TeV. 

The mass spectrum of the 4D KK-gravitons tower is more spaced  than in the RS case, as the masses for $n>0$ are given by: 
\begin{equation}
 m_n^2 = k^2 + \frac{n^2}{r_c^2}.
    \label{eq:gravitonCWmasses}
\end{equation}
The effective interaction between the particles located at IR-brane and the KK-gravitons can be written as
\begin{equation}
    \mathcal{L} = - \sum_{n=1}^\infty \frac{1}{\Lambda_n} h^n_{\mu \nu}(x)T^{\mu \nu}(x).
    \label{eq:lagrangianoCW}
\end{equation}
Note that in these models the effective interaction scale, $\Lambda_n$, depends on the mass of the KK-graviton hence each graviton couples differently to the brane particles
\begin{equation}
    \frac{1}{\Lambda_n} = \frac{1}{\sqrt{M_5^3 \pi r_c}} \left( 1 - \frac{k^2}{m_n^2} \right)^{1/2},
    \label{eq:escalaCW}
\end{equation}
except the zero mode that, as in RS framework, couples with $\Lambda_0 = \bar{M}_p$.

In order to stabilise the size of the extra-dimension, $r_c$, in the CW/LD is common to place a dilaton field in the bulk. The spectrum of the dilaton's KK-tower is given by 
\begin{align}
    m_r &= m_{\Phi_0} = \frac{9}{8}k^2 \nonumber
    \\
    m_{\Phi_n} &= k^2 + \frac{n^2}{r_c^2},
    \label{eq:escalaCW}
\end{align}
and the zero mode is identified as the radion field.

\subsubsection{Other models: Unparticles and other confining sectors}

We have described three scenarios with towers of resonances, which in some limits in the parameter space would lead to an almost continuum of broad/nearby resonances. In the next section~\ref{sec:pheno} we will discuss what this parameter space is. But before moving onto the phenomenological aspects, we will discuss other theories which could mimic the phenomenology of Warped Extra-Dimensions.

In the context of String Theory there is a huge body of work describing the idea that theories with weak gravity can be dual to strong gauge theories in a lower dimensional space, inspired by the so-called AdS/CFT~\cite{Maldacena:1997re}. Indeed, those dualities between two types of theories seem to extend beyond the specific example of the AdS/CFT set-up. To some qualitative level, one can understand models like RS as duals to a 4D strongly coupled theory, where the KK-tower is interpreted as bound states of preon quark and gluons with increasing mass but same quantum numbers. In QCD we do observe to a certain extent a tower of resonances for each $J^{CP}$ choice, for example in the vector sector we have observed the $\rho$ meson, higher up in mass the $\rho'$, followed by a continuum of broader $\rho^n$. The $J^{CP}$ properties of $\rho$, $\rho'$ and the continuum can be derived from the angular distributions of the final states. 

In this context the KK-graviton in RS is no exception: it can be interpreted as Pauli-Fierz  spin-two bound state of new quark and gluon preons. Indeed, it has been shown that the collider properties of the KK-graviton are indistinguishible from a hypothesised spin-two field from a new strongly-coupled sector~\cite{Fok:2012zk}.

But there are other possible alternatives to interpret this continuum, all closely related by dualities. For example, in Ref.~\cite{Dvali:2006az} it has been proposed that Black Hole dynamics can lead to spin-two hair, and that for low-energy strong gravity models these spin-two {\it states} could be produced at colliders~\cite{Dvali:2007wp}. 

Yet another alternative interpretation of a continuous spectrum is provided by the so-called {\it unparticles}~\cite{Georgi:2007ek}. These theories are based on the observation that a quasi-conformal sector coupled to the SM would show up at low energies as a continuous emission of energy, producing then a non-resonant tail on some distributions of SM particles.

To summarise, any interpretation of non-resonant phenomena in terms of RS parameters, or Deconstructed, or ClockWork, could be recasted as certain models of strongly-coupled gauge theories, quantum Black Hole production, or a new quasi-conformal sector in Nature.

 \section{\label{sec:pheno} Phenomenological signatures of non-resonant models}
 
 We now move onto the phenomenological description of the models described in the previous section.
 
 For the EFT case, the interpretation procedure is quite straightforward and already in the experiment's road-map. There is a classification of  possible deviations from the SM which could lead to excesses in tails, at ${\cal O}(p^2/M^2)$~\cite{Grzadkowski:2010es} and to some extent at ${\cal O}(p^4/M^4)$~\cite{Hays:2018zze,Murphy:2020rsh,Li:2020gnx}. 
 
 For the pseudo-Goldstone case, only recently the non-resonant opportunities have been explored in Ref.~\cite{Gavela:2019cmq}, where it was noted that for light particles whose resonant region is not accessible due to selection cuts, tails may be the only way to discover those states. 
 
Regarding broad/nearby resonances, in Ref.~\cite{Giudice:2017fmj} a description of non-resonant diphoton analysis was done in the context of ClockWork/Linear Dilaton.
 
 In this section we are going to explore under what circumstances our scenarios would lead to non-resonant phenomena. Our aim is to provide a simplified framework to interpret these tails in terms of the three scenarios: heavy, light and broad. For concreteness, we will focus on a simple final state, the dijet, but the conclusions can be generalized to other final states. We will start with a parton-level discussion to move onto a more realistic hadron collider simulation.
 
 \subsection{Parton-level cross sections} 
 
 \subsubsection{The heavy and light cases}
 
In the partonic cross-section ($\hat \sigma$), EFT effects would primarily appear as powers of $\hat s/M^2$, where $M$ is the scale of new heavy resonances, namely through the interference of EFT with SM amplitudes
\begin{equation}
\frac{\hat \sigma-\hat \sigma_{SM}}{\hat \sigma_{SM}} \propto \bar{c} \,  \frac{\hat s}{M^2}
\end{equation}
 which leads to a growing amplitude with the energy of the event. For the dijet case, there are a number of dimension-6 operators which could modify this cross section, notably the pure gauge $c_{3G} \, G_{\mu}^{\alpha}\, G_{\alpha}^{\beta}\, G_{\beta}^{\mu}$~\footnote{See  Ref.~\cite{Goldouzian:2020wdq} for a recent discussion on the phenomenology of this operator and Ref.~\cite{Dixon:1993xd} for a canonical reference.}. 
 
 On the other hand, if the EFT operator is CP-odd, e.g.  $\tilde c_{3G} \, G_{\mu}^{\alpha}\, G_{\alpha}^{\beta}\, \tilde G_{\beta}^{\mu}$, the partonic cross-section would be further suppressed as the leading contribution would come from the EFT$^2$ terms~\cite{Ferreira:2016jea}
 \begin{equation}
    \frac{\hat \sigma-\hat \sigma_{SM}}{\hat \sigma_{SM}} \propto \tilde c^2 \frac{\hat s^2}{M^4} \ , 
 \end{equation}
 which is steeper with $\hat s$, and conversely more suppressed with the scale of the heavy resonances $M$.
 
 Interestingly, for a light pseudo-Goldstone boson, an axion-like particle, the off-shell production is quite similar to square EFT effects~\cite{Gavela:2019cmq}. In the off-shell region, the partonic cross section looks like,
 \begin{equation}
\frac{\hat \sigma-\hat \sigma_{SM}}{\hat \sigma_{SM}} \propto   \frac{\hat s^2}{M^4} \ ,
\end{equation}
so the growth with energy is also  $E^4$ relative to the SM.

Hence one can relate the EFT quadratic effects and axion-like parameters as follows
\begin{equation}
    f_a \simeq M/\sqrt{\tilde c} \, \, \,\textrm{  \bf (ALP}  \leftrightarrow \textrm{\bf EFT)} \ ,
\end{equation}
namely, the decay constant of a light pseudo-Goldstone boson ($m_a \ll f_a$) plays the role of the scale of new heavy states and their coupling to SM particles. 

\subsubsection{The broad case: \\ matching of Warped, ClockWork and Deconstructed models}

The LHC has placed strong limits on resonances coupled to SM particles, often in the multi-TeV range. But imagine that these resonances were not isolated, but came in a tower of multiple, close-by resonances, forming an almost continuum. In that case, usual bump searches would fail {\it but} also broad resonance searches, where broad often means $\Gamma/m > 0.1$ and do not capture a quasi-continuum. In this case, only searches for excesses in tails could unveil their presence. 

In the previous section we have discussed many models which predict a spectrum with towers of states with the same quantum numbers. Here we are going to discuss which regions of their parameter space brings these spaced towers close into a continuum. We will also obtain a dictionary among the three standard interpretations we discussed: Warped Extra-Dimensions, ClockWork and Deconstruction.

  First of all, it is important to identify the free independent parameters of these three models. In the RS and CW/LD the fundamental free parameters are the curvature along the five dimension $k$ and the size $R_c$. In Deconstruction, they are the number of sites ($N$) and the scale times coupling of the interactions ($g v$). But the fundamental parameters are not the best option in order to study the phenomenology, as we show in Table \ref{tab:parameters}, but the mass gap and coupling to SM particles.

\begin{table}[]
\begin{center}
\begin{tabular}{c|c|c|c|}
\cline{2-4}
\multicolumn{1}{l|}{}                                                                                         & \multicolumn{1}{l|}{$\, \, \, \, \, $RS $\, \, \, \, \, $} & \multicolumn{1}{l|}{CW/LD} & \multicolumn{1}{l|}{Decon.} \\ \hline
\multicolumn{1}{|c|}{\multirow{2}{*}{\begin{tabular}[c]{@{}c@{}}Fundamental \\ parameters\end{tabular}}} & $k_{RS}$ & $k_{CW}$ & $N$ \\ 
\multicolumn{1}{|c|}{} & $R_{RS}$ & $R_{RS}$ & $gv$\\ \hline
\multicolumn{1}{|c|}{\multirow{2}{*}{\begin{tabular}[c]{@{}c@{}}Useful\\ parameters\end{tabular}} }&  $m_1$  & $k_{CW}$  & $M_1$ \\
\multicolumn{1}{|c|}{} & $\Lambda_{RS}$  & $M_{5 CW}$  & $gv$ \\ \hline
\end{tabular}
\end{center}
\caption{Free parameters of Warped Extra-Dimensions (RS), ClockWork/Linear Dilaton and Deconstruction. The first parameter set represents the fundamental parameters of each model. The second parameter set shows the parameters that are useful for phenomenology (mass gap and coupling).}
\label{tab:parameters}
\end{table}
 
Despite their differences,  one should be able to match these three models in the non-resonant limit, as the continumm is roughtly characterised by two parameters, the threshold (mass gap) and the height of the spectrum.  

We can start our matching exercise with  the RS and CW/LD models. In RS framework the mass gap between two consecutive resonances can be found using Eq.~\ref{eq:masas}:
 \begin{equation}
     \Delta m^{RS} \approx \frac{m_1\pi}{x_1},
     \label{deltam_RS}
 \end{equation}
 where $m_1$ is the first graviton mass and we are assuming $n \gg 1$. In the CW/LD case the masses are given by Eq.~\ref{eq:gravitonCWmasses}, leading to  
  \begin{equation}
     \Delta m^{CW} \approx \frac{1}{R_c},
     \label{deltam_CW}
 \end{equation}
 in the $k << \left(\frac{n}{r_c}\right)^2$ limit. In CW/LD model each graviton couple different to the energy-momentum tensor, Eq.~\ref{eq:escalaCW}. In the $m_n >> m_1$ limit (where the spectrum would approximate a continuum) the expression reduces to:
   \begin{equation}
      \Lambda_{CW} \approx M_5^{3/2} \sqrt{\pi r_c}.
 \end{equation}
 
 The matching between the two spectra can be obtained by equating Eq.~\ref{deltam_RS} and Eq.~\ref{deltam_CW}, which leads to the following relations
    \begin{equation}
     R_{CW} \approx \frac{x_1}{\pi m_1} \, \, \,\textrm{  \bf (CW/LD}  \leftrightarrow \textrm{\bf RS)} \ ,
 \end{equation}
 and
    \begin{equation}
      M_{5_{CW}} = \left(\frac{\Lambda_{RS}}{\sqrt{\pi R_{CW}}}\right) \, \, \,\textrm{  \bf (CW/LD}  \leftrightarrow \textrm{\bf RS)} \ ,
 \end{equation}
Now using Eq.~\ref{eq:gravitonCWmasses}, one can obtain the equivalent $k_{CW}$ value. 
 \begin{figure}[htbp]
\centering
\includegraphics[width=70mm]{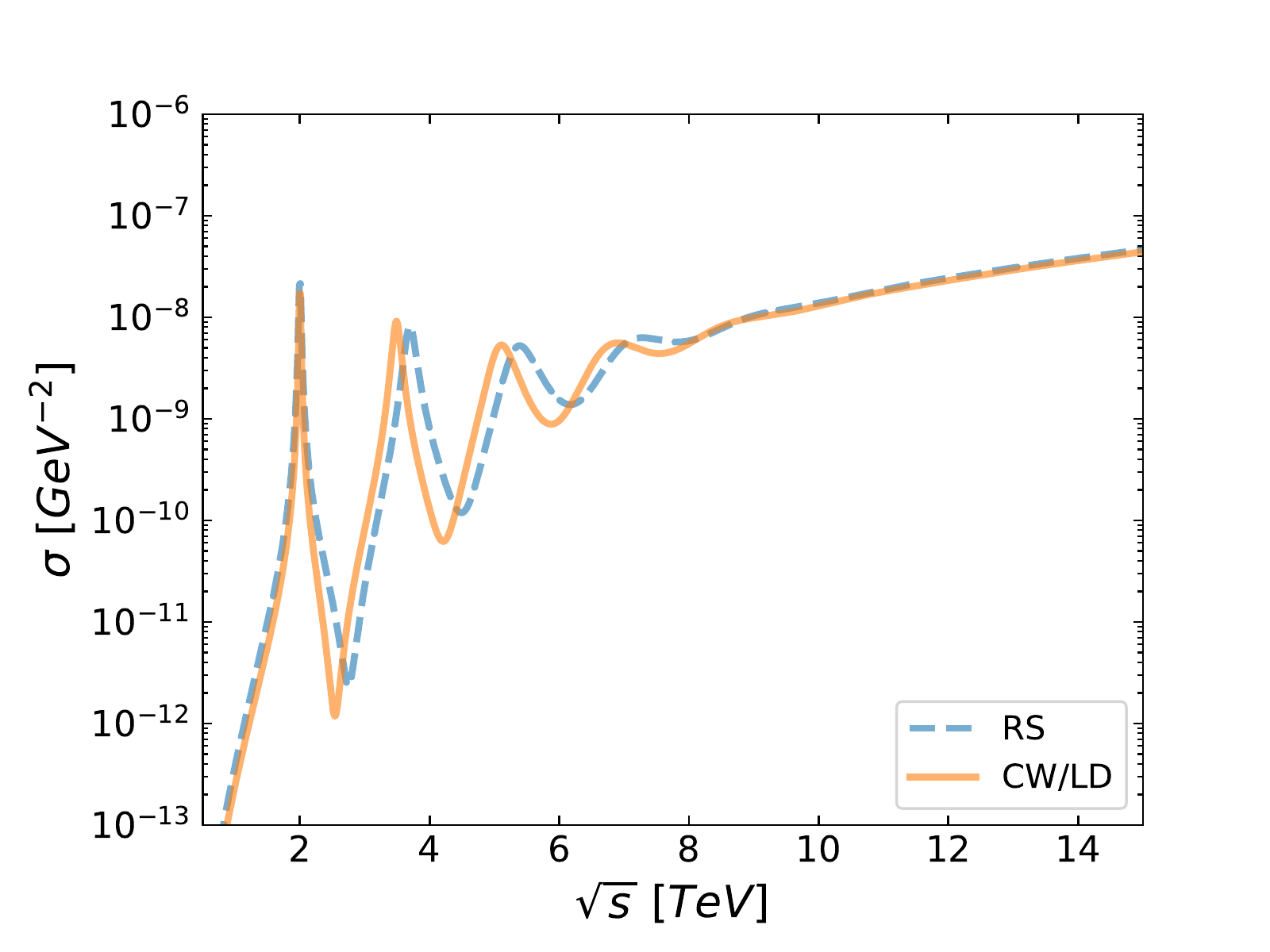}
\caption{Comparison between the RS (dashed-blue line) and the CW (solid-orange line) models in the matching region. For this plot we chose $m_1 = 2000$ GeV and $\Lambda_{RS} = 5100$ GeV which implies $M_{5_{CW}} \approx 2390$ GeV and $k_{CW} \approx 1131$ GeV. }
 \label{fig:comparacion}
\end{figure}

In Fig.\ref{fig:comparacion} we can see an example of bringing both theories to the broad/nearby resonance limit and performing this matching. The interactions of KK-resonances to SM particles are of the form given in Eq.~\ref{eq:lagrangiano}, hence grow with energy as 
\begin{equation}
    \hat \sigma \propto \frac{\hat s}{\Lambda_{RS}^2} \ .
\end{equation}

We can also bring the continuum spectrum closer to the first resonance. This situation occurs when $\Delta m = \lambda^2 \times \Gamma_1$ (where $\lambda$ is $\mathcal{O}(1)$ parameter). In this particular region the width of the resonances is more or less the value of the gap, and we obtain an overlapping between the resonances. The total decay of the first graviton is
\begin{equation}
\Gamma_n = \frac{m_n^3}{\pi\Lambda^2}(\frac{73}{240} + NP),   
\end{equation}
where NP represent the new physics contributions (possible decays to non-SM final states). Assuming this information we can write the condition for broadness as
\begin{equation}
    \Lambda_{RS} = \lambda \frac{m_1}{\pi}\sqrt{\left(\frac{73}{240} + 1\right) x_1}
    \label{condicion_continuo}
\end{equation}.

 \begin{figure}[htbp]
\centering
\includegraphics[width=70mm]{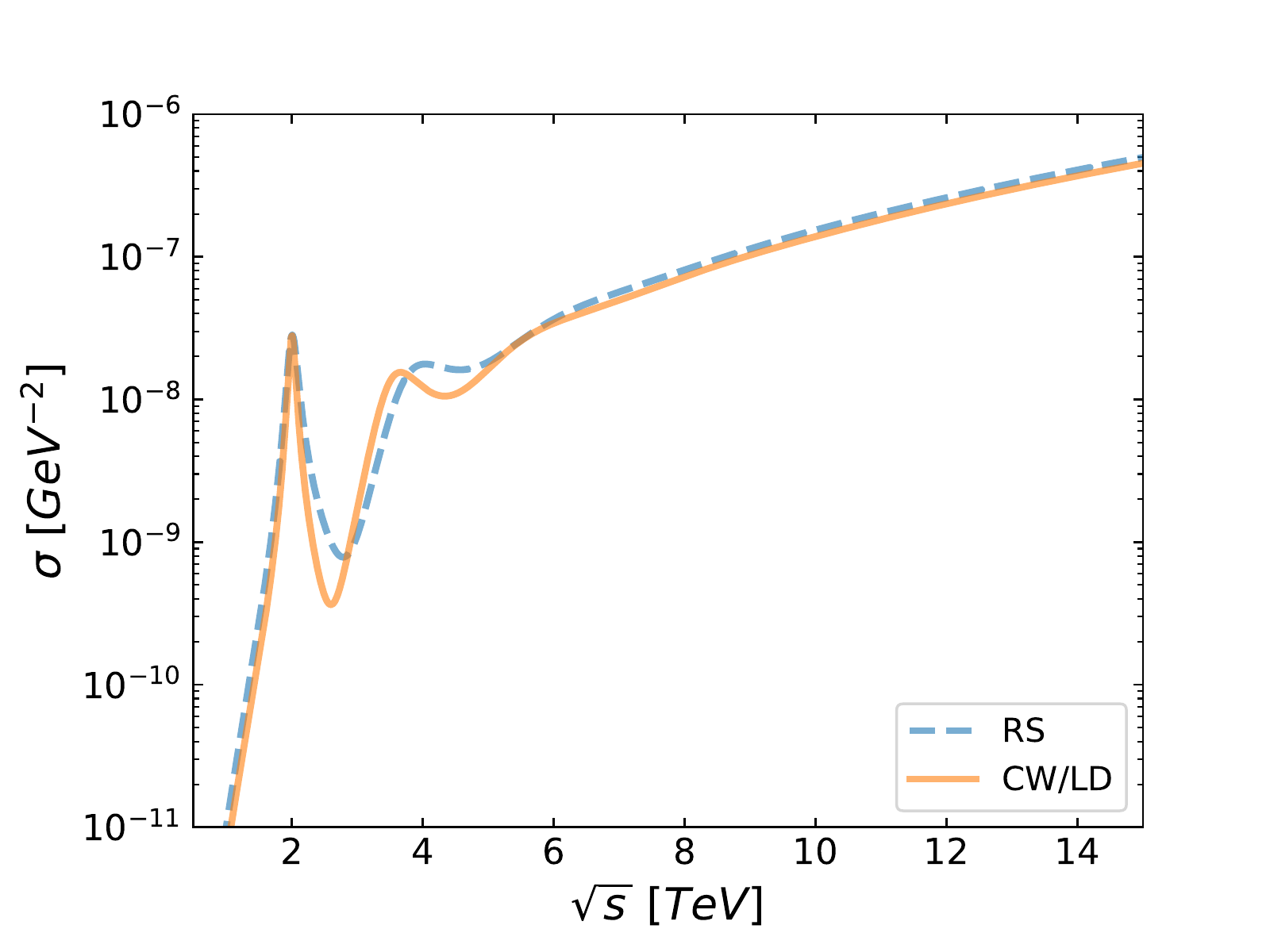}
\caption{Comparison between the RS (dashed-blue line) and the CW (solid-orange line). For this plot we choose $\lambda = 1.5$ and $NP = 1$, i.e.  $m_1 = 2000$ GeV, $\Lambda_{RS} = 2128$ GeV,  $M_{5_{CW}} \approx 1334$ GeV and $k_{CW} \approx 1131$ GeV.}
 \label{fig:comparacion}
\end{figure}

In Fig.\ref{fig:comparacion} we show an example for this kind of  spectrum, where we considered $NP = 1$ (the same order than SM contribution) while $\lambda = 1.5$, chosen so that there is no conflict with the $\Lambda_{RS}>m_1$ condition.

We have shown the relation between RS and CW/LD in the non-resonant regime. The same exercise can be done for Deconstruction models. One can easily find that the Deconstruction fundamental parameters are related to RS parameters as follows
\begin{eqnarray}
    gv &=& \Lambda_{RS} \, \, \,\textrm{  \bf (Decon.}  \leftrightarrow \textrm{\bf RS)} \ , \nonumber \\
    N + 1 &=& \frac{x_1 \Lambda_{RS}^2}{\bar{M}_{pl} \, m_1} \, \text{ln}(\bar{M}_{pl}/\Lambda_{RS})  \ .
\end{eqnarray}

 \subsection{Non-resonant tails at the Large Hadron Collider}
 
 In the previous section we have discussed the differences among the three scenarios at parton level:
 \begin{itemize}
     \item EFT, with interference with SM: grows as $\hat s/M^2$,
     \item off-shell light axion-like: grows as $\hat s^2/f_a^4$,
     \item broad/nearby resonances: quick increase from the mass gap location, grows as $\hat s/\Lambda^2$ after that. 
 \end{itemize}
 
 \begin{figure}[h!]
    \centering
    \includegraphics[scale=0.25]{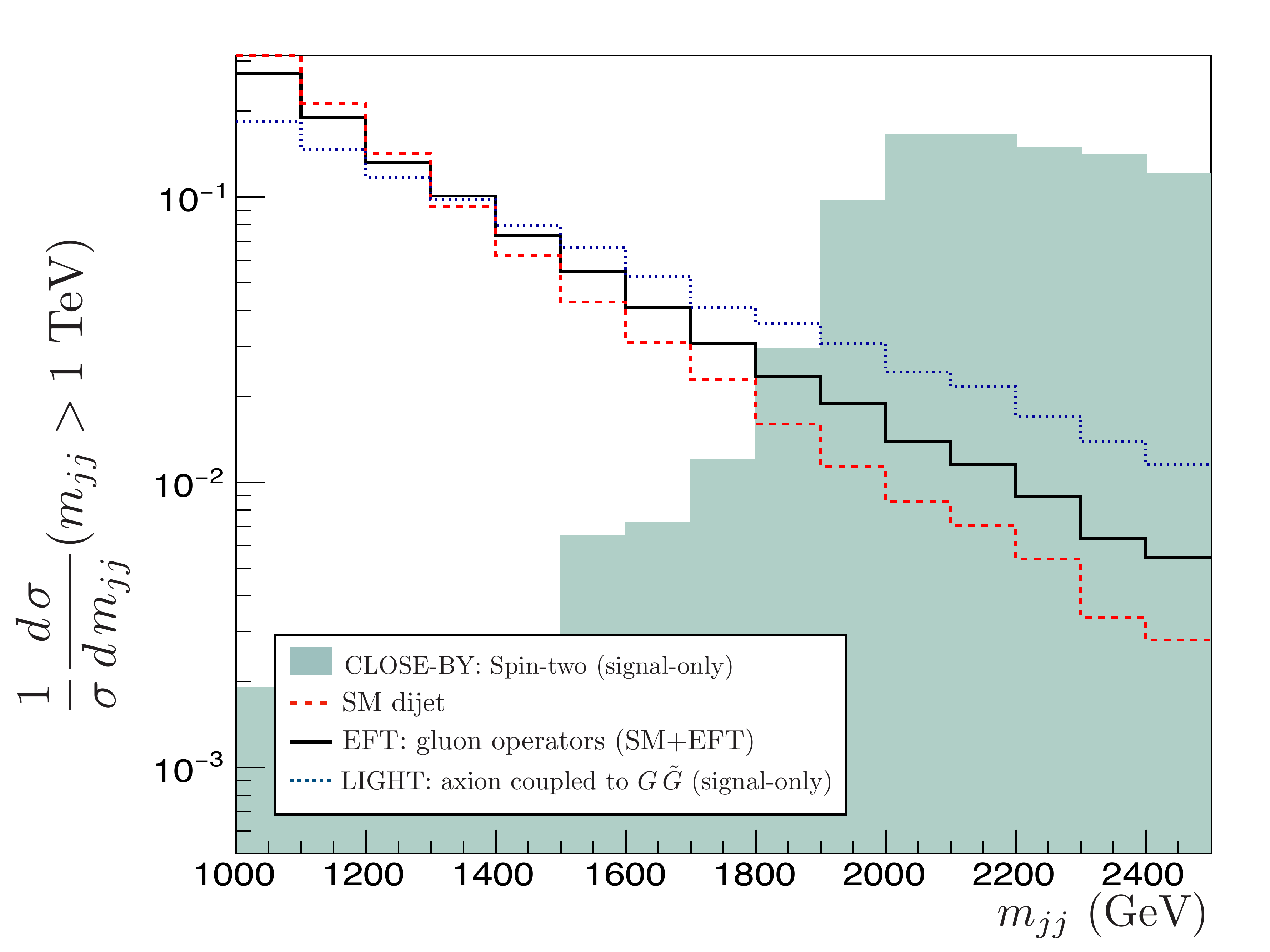}
    \caption{The $m_{jj}$ differential distribution in the SM, EFT, axion-like and broad resonance cases.}
    \label{fig:hadron}
\end{figure}

 Now we compute numerically the total cross section (including hadronization and PDF effects) at LHC energies. The results are shown in Fig.~\ref{fig:hadron}, where one can observe the hierarchies we just described: the strongest slope comes from the off-shell axion-like $\hat s^2$ behaviour, followed by the EFT $\hat s$. The broad resonance case overcomes the mass gap to follow with the $\hat s$ slope of dimension-five interactions. We also plot the SM dijet production for comparison. 
 
 We model the tower of broad resonances with a simple modification of an existing  Monte-Carlo model provided in Ref.~\cite{Christensen:2008py} and based on the implementation of spin-two resonances in Ref.~\cite{Hagiwara:2008jb}. The resonances start at 2 TeV and are set to be nearby, $\Delta m_{ij} \sim \Gamma_{i,j}$. We have set the coupling of all resonances to be the same, as in the RS models. As we are plotting signal-only normalized to the total cross-section, the specific value of $\Lambda_{RS}$ is not relevant.
 
The EFT example has been computed using the Monte Carlo implementation in Ref.~\cite{Alloul:2013naa} and with operators enabling new four- and three-gluon vertices such as  
\begin{equation}
 \frac{g_s^3 \, \bar c_{3G}}{m_W^2} \, f_{abc}   G_{\mu \nu}^{a} \, G^{b, \nu}_\rho \, G^{c,\rho \mu}
\end{equation}
where we set $\bar c_{3G}=0.1$ and plotted the SM+EFT effect, including the dominant interference.
 
 For the axion-like case, we used the Monte Carlo implementation from the study in Ref.~\cite{Brivio:2017ije} to compute the off-shell axion exchange. As for the RS model, the specific value of $f_a$ is not relevant for this normalized distribution.
 

 \section{Outlook}\label{sec:conclusions}
 In the current state of affairs at the LHC, the EFT framework is gaining support as a benchmark for interpretation of non-resonant phenomena. Within this framework, traditional searches for new physics, resonant searches, are replaced by searches for extended excesses, typically in the high-$p_T$ tails of SM distributions. 
 
 But the EFT is by no means the only way to think of non-resonant behaviour. Many scenarios, at least in some area of their pararameter space, can only be explored in the non-resonant kinematic regions. 
 
 In this paper we have described many of these type of models, classify them in terms of heavy, light or broad/nearby particles, and given a dictionary among them. For example, we have related parameters of axion-like models with EFT effects, and Warped Extra-dimensions with ClockWork and Deconstruction models.
 
 Our aim was to provide a framework for re-interpretation of non-resonant limits and for future characterisation of a possible confirmed excess. Our theory examples represent a diverse set of ideas in our area, and include axion-like particles, extra-dimensional KK-resonances, unparticles, deconstruction, clockwork, quantum black holes, and strong-coupling gauge duals of gravity theories. 
 
 We have discussed the parton- and hadron-level behaviour of dijet distributions, but all these models could be explored with the same detailed study of kinematic distributions of various final states, including  missing energy+X, massive diboson, Higgs+Z, and di-Higgs. 
 
 We hope that this work will motivate a broader view of the physics potential of collider  non-resonant searches.

\bibliography{apssamp}

\end{document}